\documentclass[a4paper]{jpconf}

\usepackage{latexsym,graphicx,epsfig}
\usepackage{amsmath,amsfonts,amssymb,amsthm}
\usepackage[
    colorlinks,%
    linkcolor=blue,citecolor=red,urlcolor=blue,
]{hyperref}

\newcommand{\R}{{\mathbb R}}

\newcommand{\maps}{\colon}

\def\stackto #1 { \, {\stackrel{#1}{\longrightarrow}}\, }
\def\stackTo #1 { {\stackrel{#1}{\Longrightarrow}} }

\newcommand{\SO}{{\rm SO}}
\newcommand{\so}{\mathfrak{so}}

\newcommand{\g}{\mathfrak{g}}

\newcommand{\define}[1]{{\bf #1}}

\newcommand{\grqc}[1]{\href{http://arxiv.org/abs/gr-qc/#1}{arXiv:gr-qc/#1}}

\newcommand{\arxiv}[1]{\href{http://arxiv.org/abs/#1/}{arXiv:#1}}

\newcommand{\webpage}[1]{{\color{blue}}\url{#1}{\color{blue}}}

\newtheorem*{unnumbered-thm}{Theorem}

\renewenvironment{proof}{\noindent
\textbf{Proof.}}{\hfill\rule{.6em}{.8em} \medskip}

\begin{document}
\title{The geometric role of symmetry breaking in gravity}

\author{Derek K.\ Wise}

\address{Institute for Theoretical Physics III, University of Erlangen--N\"urnberg, Staudtstr.~7/B2, 91054 Erlangen, Germany}

\ead{wise@theorie3.physik.uni-erlangen.de}

\begin{abstract}
In gravity, breaking symmetry from a group $G$ to a group $H$ plays the role of describing geometry in relation to the geometry the homogeneous space $G/H$.  The deep reason for this is Cartan's `method of equivalence,' giving, in particular, an exact correspondence between metrics and Cartan connections.  I argue that broken symmetry is thus implicit in any gravity theory, for purely geometric reasons. As an application, I explain how this kind of thinking gives a new approach to Hamiltonian gravity in which an observer field  spontaneously breaks Lorentz symmetry and gives a Cartan connection on space. 
\end{abstract}

The success of spontaneous symmetry breaking in condensed matter and particle physics is famous.  It explains second order phase transitions, superconductivity, the origin of mass via the Higgs mechanism, why there must be at least three generations of quarks, 
and so on.  These applications are by now standard material for  modern textbooks.

Much less famous is this: broken symmetry {\em links the geometry of gauge fields to the geometry of spacetime}.   This, in my view, is the main role of symmetry breaking in gravity. 

An early clue came in 1977, when MacDowell and Mansouri wrote down an action for general relativity using a connection for the (anti-) de~Sitter group, but invariant only under the Lorentz group  \cite{macdmans}. 
Though their work was surely inspired by spontaneous symmetry breaking, it was Stelle and West \cite{stellewest} who first made their action fully gauge invariant, breaking the symmetry dynamically using a field $y$ locally valued in (anti-) de~Sitter space.  

Whether one breaks the symmetry dynamically or `by hand,' the broken symmetry of the MacDowell--Mansouri connection plays the geometric role of relating spacetime geometry to the geometry of de Sitter space.  This is best understood using {\em Cartan geometry}, a generalization of Riemannian geometry originating in the work of \'Elie Cartan, in which the geometry of tangent spaces is generalized---in this case, they become copies of de Sitter space \cite{wise-mmgravity}.  But to explain how this works, and how symmetry breaking is involved, it helps to back up further.
  
In geometry, inklings of spontaneous symmetry breaking date from at least 1872, in the work of Felix Klein.  Ironically, to study a homogeneous space $Y$, with symmetry group $G$, one first {\em breaks} its perfect symmetry, artificially giving special significance to some point $y\in Y$.  This gives an isomorphism $Y \cong G/G_y$ as $G$-spaces, where $G_y$ is the stabilizer of $y$, allowing algebraic study of the geometry.  While $Y$ itself has $G$ symmetry, this {\em description} of it is only invariant under the subgroup $G_y$.   Different algebraic descriptions of $Y$ are however related in a $G$-equivariant way, since $G_{gy} = gG_y g^{-1}$.

This is strikingly similar to spontaneous symmetry breaking in physics.  There, one really has a family of minimum-energy states, related in a $G$-equivariant way under the original gauge group $G$.  Singling out any particular state $|0\rangle$ as `the' vacuum breaks symmetry to $G_{|0\rangle}$.  

Cartan took Klein's ideas a dramatic step further, getting an algebraic description of the geometry of a {\em non}homogeneous manifold $M$, by relating it `infinitesimally' to one of Klein's geometries $Y$.   Just as Klein geometry uses broken symmetry to get an isomorphism $Y\to G/G_y$, in Cartan geometry, the broken symmetry in a $G$ connection induces an isomorphism $e\maps T_xM \to \g/\g_y$ for each tangent space.  This is just the coframe field, also called the \define{soldering form} since identifying $T_xM$ with $\g/\g_y\cong T_yY$ effectively solders a copy of $Y$ to $M$, at each point $x$.  These copies of $Y$ are then related via holonomy of the Cartan connection, which can be viewed as describing `rolling $Y$ along $M$ without slipping' \cite{wise-mmgravity} (see also \cite[Appendix B]{sharpe}).  

Physics history unfortunately skips over Cartan geometry.  The Levi-Civita connection is adequate for the standard metric formulation of general relativity, and more general kinds of connections played no vital role in physics until some time later.  When these eventually were introduced in Yang--Mill theory, they served a purpose far removed from spacetime geometry.  Yang--Mills gauge fields are really just the principal connections of Ehresmann, who, building on Cartan's ideas, liberated connections from their bondage to classical geometry.  Ehresmann's definition, which lacks the crucial `broken symmetry' in Cartan's original version, has just the flexibility needed for gauge fields in particle physics, which are concerned only with the geometry of an abstract `internal space'---a bundle over spacetime, rather  than spacetime itself.  
On the other hand, Cartan's original version is better when it comes to studying gravity.

Concretely, a Cartan geometry may be thought of as a connection on a principal bundle (with Ehresmann's now standard definition) {\em together with} a section $y$ of the associated $Y$ bundle.    As an example, let us write a version of the MacDowell--Mansouri action, using de~Sitter space $Y\cong G/H = \SO(4,1)/\SO(3,1)$ as the corresponding Klein geometry:
\[
  I[A,y]  = \int \tr(F_y \wedge {\star}_y   F_y)
\]
The Cartan connection $(A,y)$ consists of an $\SO(4,1)$ connection $A$ and a locally de Sitter-valued field $y$.  $F$ is the curvature of $A$, calculated by the usual formula, and $F_y$ is its $\g_y$-valued part, where $\g_y \cong \so(3,1) \cong \Lambda^2\R^{3,1}$ has Hodge star operator $\star_y$.     

I have described additional examples of Cartan-geometric formulations of various gravity theories elsewhere \cite{wisesigma}, and there are many more.  But besides the diversity of specific examples, there are deep reasons that gravity, or any related `gauge theory of geometry,' {\em should} be framed in the language of Cartan geometry.   This is the subject of geometric `equivalence theorems.'  

In fact, if one believes semi-Riemannian metrics are fundamental in classical gravity, one is forced to accept Cartan connections as equally fundamental.  The reason for this is Cartan's \define{method of equivalence}, a process for proving that specified kinds of `raw geometric data' are equivalent to corresponding types of Cartan geometry \cite{gardner}.   
In the case of Riemannian geometry, solving the `equivalence problem' leads to the following theorem:  
\begin{unnumbered-thm}
A Riemannian metric determines a unique torsion-free Cartan geometry modeled on Euclidean space; conversely, a torsion-free Cartan geometry modeled on Euclidean space determines a Riemannian metric up to overall scale (on each connected component).
\end{unnumbered-thm}
\begin{proof} See Sharpe's book \cite{sharpe}.
\end{proof}  

\noindent Physically, the `overall scale' in the converse just represents a choice of length unit.  One can also show that deformed versions (or `mutations') of Euclidean geometry, namely hyperbolic and spherical geometry, lead to Cartan geometries that carry the same information \cite{sharpe}.   The Lorentzian analogs of these results are the real reason de Sitter and anti de Sitter geometries work in MacDowell--Mansouri gravity.

Riemannian geometry is but one application of the equivalence method.  There are analogous theorems, for example, in conformal geometry or Weyl geometry, relating various types of conformal structures to Cartan geometries that take the model $Y$ to be an appropriate kind of homogeneous conformal model.  Sharpe's book \cite{sharpe} contains some such theorems, and some significant work has been done on applications of {\em conformal} Cartan geometry---which often goes by the name `tractor calculus'---in physics. (See, e.g.\ \cite{waldron} and references therein.)

For now, I just want to describe one more application of Cartan geometric thinking in gravitational theory.  Besides spacetime geometry, one can also use Cartan's ideas to describe the geometry of {\em space}.  

Wheeler's term `{geometrodynamics}' originally referred to the of evolution of spatial geometries in the metric sense.  This has sometimes been contrasted with `connection dynamics' \cite{conn-dynamics}.    In light of the above equivalence theorem, however, there seems little point in establishing any technical distinction between geometrodynamics and connection dynamics, at least if we mean connections in the Cartan-geometric sense.   
The metric and connection pictures have their own advantages, but the equivalence theorem suggests we {\em should} be able to translate {exactly} between the two. 

In recent work with Steffen Gielen \cite{gielenwise}, we take an explicitly Cartan-geometric approach to evolving spatial geometries.   In this case, the symmetry breaking field $y$ lives in 3d hyperbolic space $\SO(3,1)/\SO(3)$, and can be interpretated as a \define{field of observers}, since the spacetime coframe field converts it into a unit timelike vector field.  This can be dualized via the metric to a unit {\em covector} field, which we might call a field of \define{co-observers}.  Just as observers determine a local time direction, co-observers determine local space directions, by taking their kernel.   Our strategy in the Hamiltonian formulation is to fix a field of co-observers---the infinitesimal analog of picking a spacetime folitation---but let the field of observers be determined  dynamically, as part of determining the metric.  

The result is a model in which the observer field plays a two part symmetry breaking role: first splitting spacetime fields into spatial and temporal parts, but then also acting as the symmetry breaking field in Cartan geometry of {\em space}.  This gives a Cartan-geometric Hamiltonian framework in which the spatial fields fit neatly and transparently into their spacetime counterparts and transform in an equivariant way under local Lorentz symmetry.  

Thanks to the equivalence theorem, this may be viewed as a concrete link between connection dynamics and geometrodynamics in the original sense.

\vskip .5em

It is conceivable that gravity descends from a more fundamental theory with larger gauge group, and so fits into the tradition of symmetry breaking in gauge theories. Such ideas are clearly worth pursuing (see, e.g.~\cite{percacci,randono}).  At the same time, we should not ignore the lesson of Cartan geometry: broken symmetry is the means to establishing exact correspondence between geometric structures living on tangent spaces on one hand and connections on the other.

\vskip .5em

I refer the interested reader to the bibliography of \cite{wisesigma} for many additional Cartan geometry references. I would like to thank John Baez, Julian Barbour, James Dolan and Andy Randono for helpful discussions, and especially Steffen Gielen for collaboration on \cite{gielenwise}.

\end{document}